\documentclass[12pt]{article}
\usepackage[latin1]{inputenc}
\usepackage[dvips]{graphicx}
\newcommand{\minrm}{{\rm min}}
\newcommand{\maxrm}{{\rm max}}
\newcommand{\frm}{{\rm f}}

\newcommand{\RSIIrm}{{\rm RS(II)}}

\newcommand{\exprm}{{\rm exp}}

\newcommand{\drm}{{\rm d}}

\newcommand{\crm}{{\rm c}}

\newcommand{\pa}{\partial}

\newcommand{\text}{\rm}

\newcommand{\ug}{ \; = \; }

\newcommand{\infi}{\infty}

\newcommand{\bb}{\begin{equation}}
\newcommand{\ee}{\end{equation}}
\newcommand{\bega}{\begin{eqnarray}}
\newcommand{\ega}{\end{eqnarray}}
\newcommand{\begae}{\begin{eqnarray*}}
\newcommand{\egae}{\end{eqnarray*}}

\newcommand{\h}{\hspace*{4ex}}
\newcommand{\dis}{\displaystyle}

\newcommand{\Om}{\Omega}
\newcommand{\om}{\omega}

\newcommand{\ove}{\overline}

\newcommand{\cent}{\centerline}
\newcommand{\vs}{\vspace*}

\begin{document}

\baselineskip 0.8cm

\begin{center}

{\large {\bf Focused X-shaped Pulses}$^{\: (\dag)}$}
\footnotetext{$^{\: (\dag)}$  Work partially supported by FAPESP
(Brazil), and by MIUR-MURST and INFN (Italy); previously
available as e-print ******. \ E-mail addresses for contacts:
mzamboni@dmo.fee.unicamp.br; \ shaarawi@aucegypt.edu; \
recami@mi.infn.it}

\end{center}

\vs{3mm}

\cent{ Michel Zamboni-Rached, }

\vs{0.1 cm}

\centerline{{\em D.M.O., Faculty of Electrical Engineering,
UNICAMP, Campinas, SP, Brasil.}}

\vs{0.4 cm}

\cent{ Amr M. Shaarawi$\; {^*}$ }

\vs{0.1 cm}

\cent{{\em The Physics Department, The American University in
Cairo, P.O.Box 2511, Cairo 11511, Egypt.}}

\vs{0.4 cm}

\cent{ Erasmo Recami }

\vs{0.1 cm}

\cent{{\em Facolt\`a di Ingegneria, Universit\`a statale di
Bergamo, Dalmine (BG), Italy;}} \cent{{\em {\rm and} \
INFN---Sezione di Milano, Milan, Italy.}}

\footnotetext{$\; {^*}$ On leave from the Department of
Engineering Physics and Mathematics, Faculty of Engineering,
Cairo University, Giza 12211, Egypt.}

\vs{0.5 cm}

\

{\bf Abstract  \ --} \  The space-time focusing of a (continuous)
succession of localized X-shaped pulses is obtained by suitably
integrating over their speed, i.e., over their axicon angle, thus
generalizing a previous (discrete) approach. First, new
Superluminal wave pulses are constructed, and then tailored in
such a wave to get them temporally focused at a chosen spatial
point, where the wavefield can reach for a short time very high
intensities.  Results of this kind may find applications in many
fields, besides electromagnetism and optics, including acoustics,
gravitation, and elementary particle
physics.\\

{\em PACS nos.}: \ 41.20.Jb ; \ 03.50.De ; \ 03.30.+p ; \
84.40.Az ; \ 42.82.Et ; \ 83.50.Vr ; \ \ 62.30.+d ; \ 43.60.+d ;
\  91.30.Fn ; \  04.30.Nk ; \  42.25.Bs ; \ 46.40.Cd ; \ 52.35.Lv
\ .\hfill\break
{\em OCIS codes\/}: \ 320.5550 ; \ 320.5540 .\\

{\em Keywords\/}: Localized solutions to Maxwell equations;
Superluminal waves; Bessel beams; Limited-diffraction pulses;
Finite-energy waves; Electromagnetic wavelets; X-shaped waves;
Electromagnetism; Microwaves; Optics; Special relativity;
Localized acoustic waves; Seismic waves; Mechanical waves;
Elementary particle physics; Gravitational waves

\newpage

{\bf 1. -- Introduction}\\

\h For many years it has been known that localized
(non-diffractive) solutions exist to the wave equation[1]. Some
localized wave solutions have peaks that travel at the speed of
light, while others are endowed with subluminal or Superluminal
velocities[2].  In more recent years, particular attention has
been paid to the Superluminal[3-7] Localized Waves (SLW), that
can have several applications such as high resolution imaging[8],
secure communications, non-diffractive pulse propagation in
material media[9-11], identification of buried objects[12], and
so on.

\h The most characteristic SLWs resulted to be the so-called
X-shaped solutions[13,3,4,6,14] (also named X-waves, in brief)
whose structure and behavior are by now well understood, and
experimentally reproduced[15-18]. Even their propagation along
waveguides has been investigated[19]. \ In addition, their finite
energy versions, with arbitrary frequencies and adjustable
bandwidths[6,10,20], have been also constructed. Moreover,
several investigations about constructing and generating
approximate X-shaped waves from finite apertures[21,22] have been
carried on.

\h In a recent paper by Shaarawi et al.[23], appeared in this
Journal, it has been introduced a space-time focusing technique
(called ``Temporal focusing"), that had recourse to
superpositions of localized X-waves {\em traveling with different
velocities}. The various pulses were designed to reach a given
spatial point $z=z_\frm$ at the same time $t=t_\frm$. In the
mentioned work[23], the resulting composite X-shaped wave was
synthesized as a {\em discrete sum} of individual X-waves. In the
present paper we generalize that focusing scheme, by going on, in
particular, to a continuous superposition of individual X-waves,
i.e., to integrals (instead of discrete sums). We are moreover
going to show how one can in general use any known Superluminal
solution, to obtain from it a large number of analytic
expressions for space-time focused waves, endowed with a very
strong intensity peak at the desired location. Finally, we shall
consider the case of the excitation of such pulses from
finite-size apertures. At variance with the source-free case, the
range over which the aperture-generated pulses can be focused is
limited by the field depth of the single X-wave components.

\

\

{\bf 2. -- The Discrete Space-time Focusing Method}\\

\h Let us first summarize the ``temporal focusing" scheme
developed in Ref.[23]. Since the velocity of the X-shaped waves
depends on their apex angle $\theta$ (also known as the axicon
angle), in the previous paper the space-time focusing was
achieved by superimposing a discrete number of X-waves,
characterized by different $\theta$ values. In this work, we'll
go on to more efficient superpositions for varying velocities
$V$, related to $\theta$ through the known[3,4] relation
$V=c/\cos\theta$. It will be shown in Section 3 that this
enhanced focusing scheme has the advantage of yielding analytic
(closed-form) expressions for the spatio-temporally focused
pulses.

\h Consider an axially symmetric
Superluminal\footnote{Superluminal waves, depending on $(z,t)$
through the combination $\zeta = z- Vt$ only, have infinity
energy. The versions with finite energy, written as
$\psi(\rho,z-Vt,z+Vt)$, can be found in Ref.[6]. Other types of
finite energy Superluminal pulses may be found in Ref.[20].} wave
pulse $\psi(\rho,z-Vt)$ in a dispersionless media, where $V>c$ is
the pulse velocity and $(\rho,\phi,z)$ are the cylindrical
co-ordinates. Pulses like these can be obtained by a
superposition of Bessel beams[3,4,24], viz.,

\bb \psi(\rho,z-Vt) \ug
\int_{-\infi}^{\infi}\,S(\om)\,J_0\left(\frac{\om}{c}\,\sin\theta\,\rho\right)\,
e^{i\frac{\om}{c}\,\cos\theta\,(z-\frac{c}{\cos\theta}\,t)} \drm
\om \label{sbb} \ee

where $\theta$ is the Bessel beam axicon angle, with
$V=c/\cos\theta$, and $S(\om)$ is the frequency spectrum. The
center of such pulses is localized at $z=Vt$: Many solutions of
this kind, as well as their finite energy version, can be found
in Refs.[20,6].

\h Suppose that we have now $N$ waves\footnote{Obviously, if
$\psi(\rho,z-Vt)$ is a solution of the wave equation, then
$\psi_n(\rho,z-V_n(t-t_n))$, with $t_n$ a constant, is also a
solution.} of the type $\psi_n(\rho,z-V_n(t-t_n))$, with different
velocities, $c<V_1<V_2<..<V_N$, and emitted\footnote{When we
speak of emission or arrival time, we refer to the peak of the
traveling pulse; e.g., the emission time is that at which the
peak results located at $z=0$.}   at (different) times $t_n$;
quantities $t_n$ being constants, while $n$=1,2,...$N$. \ The
center of each pulse is localized at

\bb z \ug V_n\,(t-t_n)  \label{center} \ee

\h To obtain a highly focused wave, we need all wave components
$\psi(\rho,z-Vt)$ to reach the given point, $z=z_\frm$, at the
same time $t=t_\frm$. On choosing $t_1=0$ for the slowest pulse
$\psi_1$, it is easily seen that the peak of this pulse reaches
the point $z=z_\frm$ at the time

\bb t_\frm  \ug \frac{z_\frm}{V_1} \label{tf} \ee

\h Combining Eqs.(\ref{center}) and (\ref{tf}), we obtain that
for each $\psi_n$ the instant of emission $t_n$ must be

\bb t_n \ug \left(\frac{1}{V_1}- \frac{1}{V_n}\right)z_\frm
\label{tn} \ee

\h Therefore, a solution of the type

\bb \Psi(\rho,z,t) \ug \sum_{n=1}^{N} A_n
\psi_n(\rho,z-V_n(t-t_n)) \label{soma} \ , \ee

where $A_n$ are constants, will represent a set of $N$ (initially
separated) Superluminal waves, which just reach the position
$z=z_\frm$ at the same time $t=t_\frm =z_\frm /V_1$.

\h The scheme described above was essentially developed in
Ref.[23] by using discrete X-waves: We have just replaced
summations over $\theta$ with summations over $V$. \ In the
remaining part of this work, we propose a generalization of that
idea, which can yield new classes of exact Superluminal
solutions, besides providing enhanced focusing effects.

\

{\bf 2.1 -- The new space-time focusing scheme}\\

\h In this Section, we extend the previous ``temporal focusing"
approach[23] by considering a continuous superposition, namely,
by integrating over the velocity (instead of a discrete sum over
the angle $\theta$).

\h Combining Eqs.(\ref{tn}) and (\ref{soma}), and going on to the
integration over $V$, one gets

\bb \Psi(\rho,z,t) \ug \int_{V_\minrm }^{V_\maxrm }\, \drm V \,
A(V) \, \psi\left(\rho,z-V\left(t-\left(\frac{1}{V_\minrm }-
\frac{1}{V}\right) z_\frm \right)\right) \label{int} \ , \ee

where $V$ is the velocity of the wave $\psi$ in Eq.(\ref{sbb}).
In the integration, $V$ is considered as a continuous variable in
the interval $[V_\minrm ,V_\maxrm ]$. In Eq.(\ref{int}), $A(V)$ is
the velocity-distribution function that specifies the
contribution of each wave component (with velocity $V$) to the
integration. The resulting wave $\Psi(\rho,z,t)$ can have a more
or less strong amplitude peak at $z=z_\frm$, at time $t_\frm =
z_\frm /V_\minrm $, depending on $A(V)$ and on the difference
$V_\maxrm - V_\minrm$. \ Let us notice that also the resulting
wavefield will propagate with a Superluminal velocity, depending
on $A(V)$ too. In the cases when $A(V)$ can be actually
considered a distribution function, namely, when $A(V)>0\,\, ,
\,\,\, \forall \ V$, and $\int_{V_\minrm }^{V_\maxrm }\, \drm V
\, A(V) < \infty$, we can heuristically expect that the mean
velocity $\ove{V}$ of the field (\ref{int}) will be

\bb \ove{V} \approx \dis{\frac{\int_{V_\minrm }^{V_\maxrm } \,
A(V)\,V\,\drm V} {\int_{V_\minrm }^{V_\maxrm }\, A(V)\,\drm V}} >
c  \ . \label{V} \ee

\h In the cases when the velocity-distribution function is well
concentrated around a certain velocity value, one can expect the
wave (\ref{int}) to increase its magnitude and spatial
localization while propagating. Finally, the pulse peak acquires
its maximum amplitude and localization (at the chosen point
$z=z_\frm$, and at time $t=z_\frm /V_\minrm $, as we know).
Afterwards, the wave suffers a progressive spreading, and a
decreasing of its amplitude.

\

\

{\bf 3. -- Enhanced Focusing Effects by Using Ordinary X-Waves}\\

\h Here, we present a specific example by integrating, in
Eq.(\ref{int}), over the standard, {\em ordinary\/}[3,4] X-waves:

\bb X(\rho,z-Vt) \ug \dis{\frac{V}{\sqrt{(aV-i\,(z-Vt))^2 +
\left(\frac{V^2}{c^2} -1 \right)\rho^2 }}} \ . \label{X} \ee

The ``classical" solution (8) can be obtained[3,6,4] by
substituting the spectrum \ $S(\om)= \Om(\om) \; \exprm(-a\,\om)$
\ into Eq.(1), where $a$ is a constant that defines the
bandwidth, $\Delta\om=1/a$, and $\Om(\om)$ is the
step-function.\footnote{The step-function, or Heaviside function,
assumes the values $\Om = 0$ for $\om < 0$ and $\Om = 1$  for
$\om \geq 0$, as is well-known.}  When using the zeroth-order
X-waves above, however, the largest spectral amplitudes are
obtained for low frequencies. For this reason, one may expect that
the solutions considered below will be suitable mainly for low
frequency applications.

\h Let us choose, then, the function $\psi$ in the integrand of
Eq.(\ref{int}) to be $\psi(\rho,z,t) \equiv X(\rho, z -
V(t-(1/V_\minrm -1/V)z_\frm))$, viz.,

\bb \psi(\rho,z,t) \equiv X \ug
\dis{\frac{V}{\sqrt{\left[aV-i\left( z - V\left( t -
\left(\frac{1} {V_\minrm }-\frac{1}{V}   \right)z_\frm \right)
\right)\right]^2 + \left(\frac{V^2}{c^2} -1 \right)\rho^2 }}}
\label{psiX} \ee


\h After some manipulations, one obtains the analytic {\em
integral solution}

\bb \Psi(\rho,z,t) \ug \int_{V_\minrm }^{V_\maxrm }\,
\dis{\frac{V\,A(V)} {\sqrt{PV^2 + QV + R}} }\drm V \label{intx}
\ee

with

\bb
\begin{array}{l}
P \ug \left[ \left( a+i\left(t-\frac{z_\frm}{V_\minrm
}\right)\right)^2
+ \frac{\rho^2}{c^2}\right]\\
\\
Q \ug 2\left(t-\frac{z_\frm}{V_\minrm } - ai\right)(z-z_\frm) \\
\\
R \ug \left[-(z-z_\frm)^2 - \rho^2 \right] \label{P}
\end{array}
\ee

\

{\bf 3.1 -- Some examples}\\

\h In what follows, we illustrate the behavior of our new
spatio-temporally focused pulses, by taking into consideration
{\em four} different velocity distributions $A(V)$.

\

\h {\em First case\/}:

Let us consider our ``integral solution" (\ref{intx}) with

\bb A(V) \ug 1 \label{a1} \ . \ee

In this case, the contribution of the X-waves is the same for all
velocities in the allowed range $[V_\minrm ,V_\maxrm]$, and
Eq.(\ref{intx}) yields

\bb \Psi(\rho,z,t) \ug \int_{V_\minrm }^{V_\maxrm }\,
\dis{\frac{V}{\sqrt{PV^2 + QV + R}} }\drm V \label{inta1} \ . \ee

On using identity 2.264.2 of Ref.[25], we get the particular
solution

\bb
\begin{array}{clcr}
\Psi(\rho,z,t) \!\!&=  \dis{\frac{\sqrt{PV_\maxrm ^2 + QV_\maxrm
+ R}
- \sqrt{PV_\minrm ^2 + QV_\minrm  + R}}{P}}\\
\\
&\;\;\; \dis{ + \frac{Q}{2P^{3/2}}\,{\rm
ln}\left(\frac{2\,\sqrt{P(PV_\minrm ^2 + QV_\minrm  + R)} +
2PV_\minrm  + Q}{2\,\sqrt{P(PV_\maxrm ^2 + QV_\maxrm  + R)} +
2PV_\maxrm  + Q}\right)} \ , \label{um}
\end{array}
\ee

where $P$, $Q$ and $R$ are given in Eq.(\ref{P}). A 3-dimensional
(3-D) plot of this function is provided in Fig.1; where we have
chosen $a=10^{-12}$ s, \ $V_\minrm =1.001\; c$, \ $V_\maxrm
=1.005\; c$ and $z_\frm =200\;$cm. It can be seen that this
solution exhibits a rather evident space-time focusing. An
initially spread-out pulse (shown for $t=0$) becomes highly
localized at $t=t_\frm=z_\frm /V_\minrm=6.66\;$ns, the pulse peak
amplitude at $z_\frm$ being $40.82$ times greater than the
initial one. In addition, at the focusing time $t_\frm$ the field
is much more localized than at any other times. The velocity of
this pulse is approximately $\ove{V}=1.003\; c$.

\

\begin{figure}[!h]
\begin{center}
 \scalebox{1.3}{\includegraphics{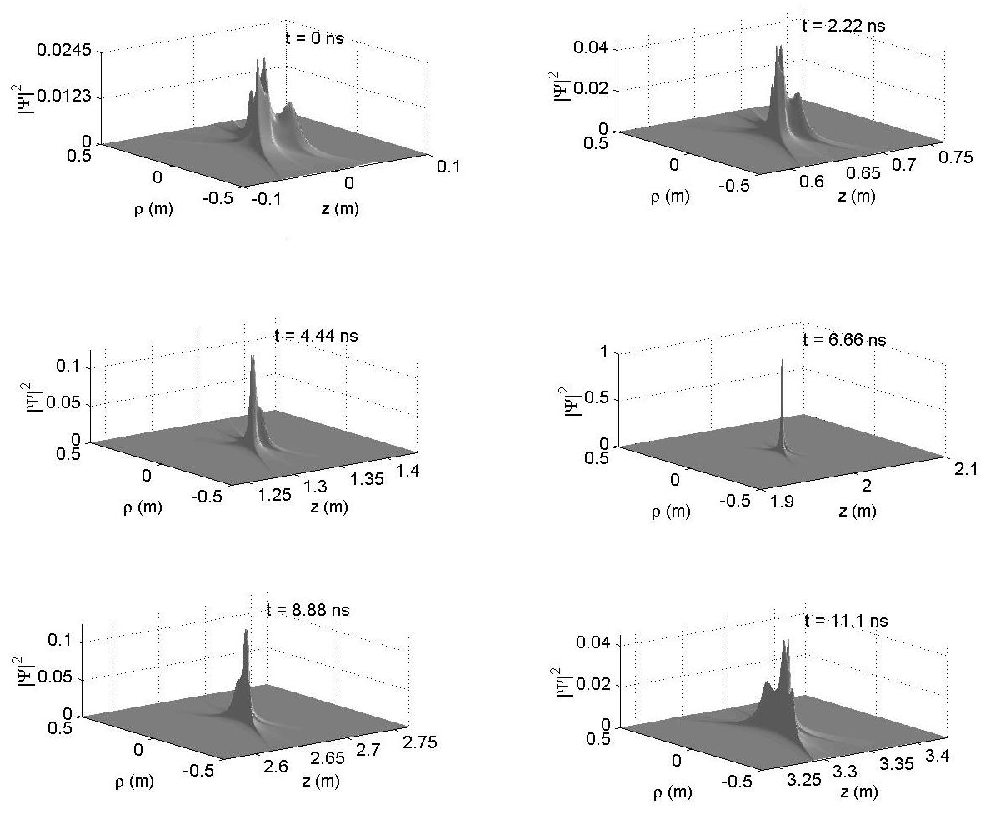}}
\end{center}
\caption{} \label{fig1}
\end{figure}

\

\newpage

\h {\em Second case\/}:

In this case we choose

\bb A(V) \ug \frac{1}{V} \ee

and Eq.(\ref{intx}) gives

\bb \Psi(\rho,z,t) \ug \int_{V_\minrm }^{V_\maxrm }\,
\dis{\frac{1}{\sqrt{PV^2 + QV + R}} }\drm V \label{intav} \ .\ee

\

On using the identity 2.261 in Ref.[25], we obtain the new
particular solution

\bb \Psi(\rho,z,t) \ug \dis{\frac{1}{\sqrt{P}}\,{\rm ln}
\left(\frac{2\,\sqrt{P(PV_\maxrm ^2 + QV_\maxrm  + R)} +
2PV_\maxrm  + Q} {2\,\sqrt{P(PV_\minrm ^2 + QV_\minrm  + R)} +
2PV_\minrm  + Q} \right)} \ . \label{dois} \ee


\

\h {\em Third case\/}:

On substituting the velocity-distribution function

\bb A(V) \ug \frac{1}{V^2} \ee

into Eq.(\ref{intx}), one gets

\bb \Psi(\rho,z,t) \ug \int_{V_\minrm }^{V_\maxrm }\,
\dis{\frac{1}{V\,\sqrt{PV^2 + QV + R}} }\drm V \label{intav2} \ee

Because of identity 2.269.1 in Ref.[25], the above integration
forwards the further particular solution

\bb \Psi(\rho,z,t) \ug \dis{\frac{1}{\sqrt{R}} \; {\rm
ln}\left(\frac{V_\maxrm (2R + QV_\minrm  + 2\sqrt{R(PV_\minrm ^2 +
QV_\minrm  + R)})}{V_\minrm (2R + QV_\maxrm  + 2\sqrt{R(PV_\maxrm
^2 + QV_\maxrm  + R)})} \right) } \ . \label{tres} \ee

\h {\em Fourth case\/}:

With the velocity-distribution function given by

\bb A(V) \ug \frac{1}{V^3} \ee

we have from Eq.(\ref{intx}):

\bb \Psi(\rho,z,t) \ug \int_{V_\minrm }^{V_\maxrm }\,
\dis{\frac{1}{V^2\,\sqrt{PV^2 + QV + R}} }\drm V \ .
\label{intav3} \ee

On using the identity 2.269.2 of Ref.[25], we get the last
particular solution

\bb
\begin{array}{clcr}
\Psi(\rho,z,t) \!\!&=  \dis{\frac{\sqrt{PV_\minrm ^2 + QV_\minrm
+
R}}{RV_\minrm } - \frac{\sqrt{PV_\maxrm ^2 + QV_\maxrm  + R}}{RV_\maxrm }}\\
\\
&\;\;\; \dis{ + \; \frac{Q}{2R^{3/2}} \; {\rm
ln}\left(\frac{V_\minrm (2\,\sqrt{R(PV_\maxrm ^2 + QV_\maxrm  +
R)} + 2R + QV_\maxrm )}{V_\maxrm (2\,\sqrt{R(PV_\minrm ^2 +
QV_\minrm  + R)} + 2R + QV_\minrm )}\right) } \ . \label{quatro}
\end{array}
\ee

\

\

{\bf 4. -- Enhanced Space-time Focusing by Using Higher-Order
X-Waves of
Arbitrary Frequencies and Adjustable Bandwidths}\\

\h The scheme presented in the preceding Section (confined to
zeroth-order X-waves) can be extended to higher-order X-waves.
Namely, one can use in the integrand of Eq.(\ref{int}) the
various time-derivatives of the ordinary X-waves, which have been
shown[26] to constitute an infinite set of generalized X-wave
solutions. This procedure will provide us with spatio-temporally
focused Superluminal pulses that can have any arbitrary
frequencies and adjustable bandwidths[6,10]. It has been also
shown[26,7,10,6] that time derivatives of the X-waves can be
constructed by substituting into Eq.(\ref{sbb}) the frequency
spectrum $S(\om)= \Om(\om) \; \om^m\,\exp (-a\,\om)$, with $m$ an
integer.
A more general approach for obtaining infinite series of X-shaped
solutions through suitable differentiations of the ordinary X-wave
can be found in Ref.[6]; whilst more details about the properties
of the frequency-spectra, which allows for closed-form solutions,
can be found in the mentioned Refs.[10,7,6].

\h Namely: by using appropriate values for the parameters $m$ and
$a$, it is possible to shift the central frequency, $\om_\crm$,
and adjust the bandwidth, $\Delta\om$, to any desired
values[6,7,10]. The relationships among $\om_\crm$, $\Delta\om$,
$a$ and $m$ are

\bb \left\{
\begin{array}{l}
\dis{\om_\crm \ug \frac{m}{a}}\\
\\
m \ug \dis{\frac{1}{\pm\frac{\Delta\om_{\pm}}{\om_\crm} - {\rm
ln}\left[1 \pm \frac{\Delta\om_{\pm}}{\om_\crm}\right]}}
\end{array} \right. \ , \label{wc}
\ee

where[7,10] $\Delta\om_+$ ($>0$) is the bandwidth to the right,
and $\Delta\om_-$ ($>0$) is the bandwidth to the left of
$\om_\crm$; \ so that \ $\Delta\om = \Delta\om_+ + \Delta\om_- $.

\h It is easy to show that[6,7,10]

\bb
\begin{array}{clcr}
X^{(m)}(\rho,z-Vt) \!\!& \dis{\equiv
\int_{0}^{\infi}\,\om^m\,e^{-a\,\om}\,J_0\left(\frac{\om}{c}\,\sin\theta\,\rho\right)\,
e^{i\frac{\om}{c}\,\cos\theta\,(z-\frac{c}{\cos\theta}\,t)}} \drm \om \\
\\
\!\!&\dis{= i^m\,\frac{\pa^m X}{\pa t^m}} \ ,
\end{array}\label{difX}
\ee

\

\h Let us now choose the function $\psi$ in the integrand of
Eq.(\ref{int}) to be \ $\psi(\rho,z,t) \equiv i^m\,X^{(m)}(\rho,
z - V(t-(1/V_\minrm -1/V)z_\frm))$; \ equation (\ref{int}), then,
yields the new (analytic) {\em integral solution}

\bb \Psi(\rho,z,t) \ug \dis i^m \frac{\pa^m}{\pa
t^m}\int_{V_\minrm }^{V_\maxrm }\, \dis{\frac{V\,A(V)}
{\sqrt{PV^2 + QV + R}} }\drm V \ , \label{intdx} \ee

with $P$, $Q$ and $R$ given by Eq.(\ref{P}).

\h Next, let us consider the cases of {\em four} different
velocity-spectra, similar to the ones used in subsection 3.1.

\

{\bf 4.1 -- Some examples}\\

\h {\em First case\/}:

Consider the integration in Eq.(\ref{intdx}) with

\bb A(V) \ug 1 \ . \label{a1} \ee

On using the result in our previous Eq.(\ref{um}), we find the
particular solution

\bb
\begin{array}{clcr}
\Psi(\rho,z,t) \!\!&= \dis{i^m\,\frac{\pa^m}{\pa
t^m}\left[\frac{\sqrt{PV_\maxrm ^2 +
QV_\maxrm  + R} - \sqrt{PV_\minrm ^2 + QV_\minrm  + R}}{P}\right.}\\
\\
&\;\;\; \left.\dis{ + \; \frac{Q}{2P^{3/2}} \; {\rm
ln}\left(\frac{2\,\sqrt{P(PV_\minrm ^2 + QV_\minrm  + R)} +
2PV_\minrm  + Q}{2\,\sqrt{P(PV_\maxrm ^2 + QV_\maxrm  + R)} +
2PV_\maxrm  + Q}\right)} \right] \ . \label{umd}
\end{array}
\ee


\

\h {\em Second case\/}:

Let us substitute

\bb A(V) \ug \frac{1}{V} \ee

into Eq.(\ref{intdx}). On using the result in Eq.(\ref{dois}),
one finds the new particular solution

\bb \Psi(\rho,z,t) \ug i^m\,\frac{\pa^m}{\pa
t^m}\left[\dis{\frac{1}{\sqrt{P}}\;{\rm ln}
\left(\frac{2\,\sqrt{P(PV_\maxrm ^2 + QV_\maxrm  + R)} +
2PV_\maxrm  + Q} {2\,\sqrt{P(PV_\minrm ^2 + QV_\minrm  + R)} +
2PV_\minrm  + Q} \right)}\right] \ . \label{doisd} \ee


\

\h {\em Third case\/}:

On considering the velocity spectrum

\bb A(V) \ug \frac{1}{V^2} \ee

we obtain, after combining Eqs.(\ref{intdx}) and (\ref{tres}),
the further particular solution

\bb \Psi(\rho,z,t) \ug i^m\,\frac{\pa^m}{\pa
t^m}\left[\dis{\frac{1}{\sqrt{R}}\; {\rm ln}\left(\frac{V_\maxrm
(2R + QV_\minrm  + 2\sqrt{R(PV_\minrm ^2 + QV_\minrm  +
R)})}{V_\minrm (2R + QV_\maxrm  + 2\sqrt{R(PV_\maxrm ^2 +
QV_\maxrm  + R)})} \right) }\right] \ . \label{tresd} \ee

The plots in Fig.2 show that this solution implies a great
space-time focusing effect.  On using $m = 1$, \  $a=10^{-12}$ s,
\ $V_\minrm =1.001\; c$, \ $V_\maxrm =1.005\; c$, and $z_\frm
=200\;$cm, the peak amplitude at $z=z_\frm$ results to be $1000$
times larger than the initial one, while, at the focusing time
$t_\frm =6.66$ns the field is much more localized than at any
other times. The velocity of this pulse is approximately
$\ove{V}=1.0029\; c$.

\begin{figure}[!h]
\begin{center}
 \scalebox{1.3}{\includegraphics{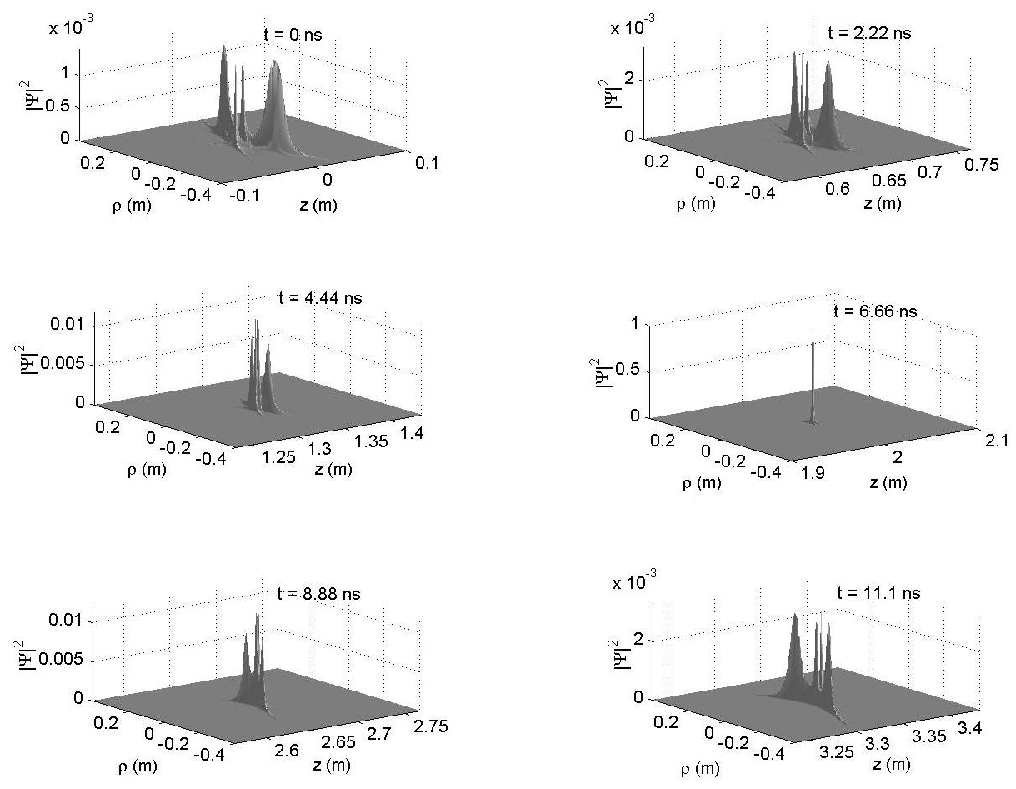}}
\end{center}
\caption{} \label{fig2}
\end{figure}

\h {\em Fourth case\/}:

On substituting the velocity spectrum

\bb A(V) \ug \frac{1}{V^3} \ee

into Eqs.(\ref{intdx}) and using the result (\ref{quatro}), we
get the last particular solution

\bb
\begin{array}{clcr}
\Psi(\rho,z,t) \!\!&= \dis{i^m\,\frac{\pa^m}{\pa t^m}}\left[
\dis{\frac{\sqrt{PV_\minrm ^2 + QV_\minrm  +
R}}{RV_\minrm } - \frac{\sqrt{PV_\maxrm ^2 + QV_\maxrm  + R}}{RV_\maxrm }}\right.\\
\\
&\;\;\; \left.\dis{ + \; \frac{Q}{2R^{3/2}}\;{\rm
ln}\left(\frac{V_\minrm (2\,\sqrt{R(PV_\maxrm ^2 + QV_\maxrm  +
R)} + 2R + QV_\maxrm )}{V_\maxrm (2\,\sqrt{R(PV_\minrm ^2 +
QV_\minrm  + R)} + 2R + QV_\minrm )}\right) }\right] \ .
\label{quatrod}
\end{array}
\ee

\

\

{\bf 5. -- A More General Formulation}\\

\h The results presented in the preceding Sections demonstrate
that in Eq.(\ref{int}) one can utilize any kind of Superluminal
pulses towards the goal of producing large space-time focusing
effects. A further generalization of our focusing scheme is
introduced in the present Section.

\h Let us recall that a Superluminal wave with axial symmetry can
be rewritten[6] as a superposition of Bessel beams expressed in
terms of $V$ (instead of $\theta$):

\bb \psi(\rho,z-Vt) \ug
\dis{\int_{-\infty}^{\infty}\,S(\om)\,J_0\left(\frac{\om}{V}\,\rho\,\sqrt{\left(\frac
{V^2}{c^2}-1
\right)}\,\,\right)\,e^{i\,\frac{\om}{V}(z-Vt)}\,\drm\om } \ ,
\label{geral1} \ee

where $J_0$ is the zeroth-order Bessel function and $S(\om)$ is
the frequency spectrum of the pulse $\psi(\rho,z-Vt)$. Now, by
using the focusing method represented by our Eq.(\ref{int}), one
gets

\bb \begin{array}{clcr} \Psi(\rho,z,t) = \dis{
\int_{V_\minrm}^{V_\maxrm} \drm V \, A(V) \,} & \dis{
\int_{-\infty}^{\infty}\,S(\om)\,J_0\left(\frac{\om}{V}\rho\sqrt{\left(\frac{V^2}{c^2}-1
\right)}\,\right) } \\
\\
& \dis{ \times \, \exp
\left[i\frac{\om}{V}\,\{z-V\,[t-\,(\frac{1}{V_\minrm}-
\frac{1}{V}\,) z_\frm \,]\,\} \right] \drm\om }\; ,
\end{array} \label{geral2} \ee
which can be rewritten as

\bb \Psi(\rho,z,t) = \dis{ \int_{V_\minrm}^{V_\maxrm} \drm V \,
\int_{-\infty}^{\infty} B(\om,V)\,J_0\left(\frac{\om}{V} \rho
\sqrt{\left(\frac{V^2}{c^2}-1 \right)}\,\right) e^{i
\frac{\om}{V}(z-Vt)} e^{i\om\left(\frac{1}{V_\minrm }-
\frac{1}{V} \right)z_\frm}\,\drm\om } \; . \label{geral3} \ee

\h From Eq.(\ref{geral3}) it can be inferred that also the
frequency-velocity weight functions of the form

\bb \ove{S}(\om,V) \, \equiv \,
B(\om,V)\,e^{i\,\om\left(\frac{1}{V_\minrm }- \frac{1}{V}
\right)z_\frm} \label{spec} \ee

are able to produce spatio-temporally focused pulses that
propagate, once more, at Superluminal speeds. Such pulses can be
considered as a continuous superposition of Bessel Beams of
different frequencies {\em and} different velocities. Moreover,
each Bessel beam, endowed with velocity $V$ and angular frequency
$\om$, possesses also a different phase with respect to the
others, given by \ ${\rm exp}(i\,\om(1/V_\minrm - 1/V)z_\frm$. \
In other words, the frequency-velocity spectra, generators of
focused Superluminal pulses, determine not only the amplitudes of
each Bessel beam in the superposition, but also the relative
phases among them. \ Along similar lines, Mugnai et al.[27]
demonstrated that one can generate {\em beams} with a very hight
optical resolving power by a superposition of Bessel beams with
suitably chosen relative phase-delays. This was achieved by the
use of a paraffin torus in each one of the coronae adopted for
producing the various Bessel beams, the required phase delays
being determined by the paraffin. \ In a sense, the
superpositions introduced in this work appear to reveal that an
analogous phenomenon holds in the case of {\em pulses} too.

\

\

{\bf 6. -- Focused Pulses Generated from Finite Apertures}\\

\h In the preceding Sections, we have demonstrated the
effectiveness of our space-time focusing scheme in the case of
source-free (composite) X-wave pulses. The analysis presented in
Section 5 can be actually regarded as a powerful tool, that may
be used to tailor an initially spread pulse, with the aim of
focusing it at a pre-chosen point in space-time. The situation
becomes more involved, however, when such a tailored pulse is
generated from an aperture having a {\em finite} radius: In fact,
the X-wave components having different velocities are
characterized by {\em different} diffraction-free lengths (or
``field depths"). More specifically, \
$z_\drm=D/(2\sqrt{\left(V/c\right)^2-1})$, \ where $\,D\,$ is the
diameter of the circular aperture.

\h To appreciate the effect of the finite size of the aperture on
the space-time focusing scheme, we calculate the field radiated
by a finite aperture using the Rayleigh-Sommerfeld (II) formula,
viz.,

\bb \Psi_{\RSIIrm} \left(\rho,z,t\right)=\int_0^{2\pi}\drm\phi'\int_0^{D/2}\drm\rho'\rho'%
\frac{1}{2\pi
R}\Biggl\{\Bigl[\Psi\Bigr]+\Bigl[\partial_{ct'}\Psi\Bigr]\frac{z-z'}{R}\Biggr%
\} \ . \label{rsii} \ee

\h The quantities enclosed by the square brackets are evaluated
at the retarded time $ct'=ct-R$. The distance
$R=\sqrt{\left(z-z'\right)^2+\rho^2+\rho'^2-2\rho\rho'\cos\left(\phi-\phi'%
\right)}$ is the separation between source and observation points.
Assuming the initial excitation to be that of Eq.(\ref{um}), we
have calculated the radiated field for the parameters
$a=10^{-12}$ s, \ $V_{min\,}=1.001\;c$ and $V_\maxrm =1.005\;c$.
The aperture radius has been chosen to equal 20 cm. For these
parameter values, the diffraction-free lengths associated with
the chosen values of $V_\minrm $ and $V_\maxrm $ are 447.1 and
199.75 cm, respectively. The crucial factor that determines the
focusing power of the tailored pulses is the focusing distance
$z_\frm$. Unlike the source-free case, the focusing distance is
influenced by the finite size of the aperture. For the chosen
parameters, all X-wave components undergo very little decay over
distances $z<199.75\;$cm. By contrast, most of the X-wave
components will be decaying at a very fast rate for
$z>447.1\;$cm. In the intermediate range $199.75 < z <
447.1\;$cm, an increasing portion of the components will decay as
the distance from the source increases.

\h To illustrate the behavior described above, we provide plots
of the axial profiles of the generated power $\big|\Psi\big|^2$
for $z_\frm =200$ and 300 cm. The first focusing point is chosen
at the edge of the diffraction-free region, while the second is
chosen in the middle of the intermediate region. The plots
displayed in Figs.3 and 4 depict the behavior of the pulse
radiated from a finite-size aperture for $z_\frm =200$ cm. The
3-D surface plots in Figs.3 show the shape of the initial
excitation on the aperture plane $z'=0$, and the shape of the
source-free pulse at the focusing point $z=z_\frm =200$ cm. In
Figs.4, we provide plots of the axial profiles of the field
radiated from the finite aperture at distances $z=50$, 150, 200
and 250 cm, respectively. One should notice that the focusing
amplitude is comparable to that of the focused source-free pulse
at $z=z_\frm$. The 3-D plot of the focused pulse radiated from
the aperture is therefore expected to be similar to the one shown
in Fig.3b for the source-free pulse. The plots in Figs.3 and 4
show, furthermore, that the peak power of the pulse is amplified
by 40 times. The fact that at $z=200$ cm the pulse radiated from
the aperture resembles the source-free pulse is a confirmation of
our qualitative prediction that, for distances $z<199.75$ cm, all
X-wave components contributing to the initial excitation of the
source are diffraction-free.

\h As a second example, we have chosen $z_\frm =300$ cm. Keeping
all other parameters equal to the ones used for Figs.3 and 4, we
find that the peak power of the aperture-radiated pulse at the
focusing point is lower than that of the source-free pulse. The
3-D plots in Figs.5 show the source free pulse at the aperture
plane and at $z=z_\frm =300$ cm. The axial profile of the pulse
generated from the finite aperture is shown in Figs.6 for $z=75$,
225, 285 and 375 cm, respectively. The peak at $z=z_\frm =300$ cm
is not shown because the peak focusing occurs, instead, at
$z=285$ cm. This is another manifestation of the fact that
contributions from certain X-wave components, constituting the
initial excitation, are lost, because such components have
surpassed their diffraction-free range. Another confirmation of
this behavior is that the peak power is amplified 10 times only,
instead of the 75 times expected for the source-free pulse.


\

\begin{figure}[!h]
\begin{center}
 \scalebox{1.3}{\includegraphics{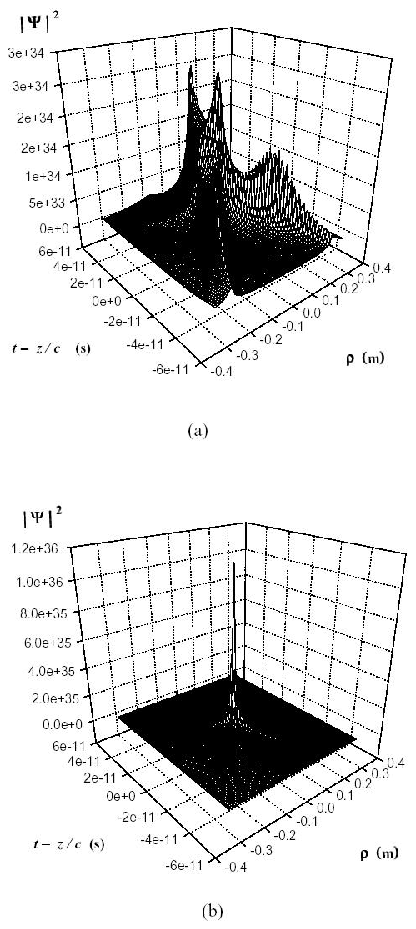}}
\end{center}
\caption{} \label{fig4}
\end{figure}

\

\newpage

\begin{figure}[!h]
\begin{center}
 \scalebox{1.4}{\includegraphics{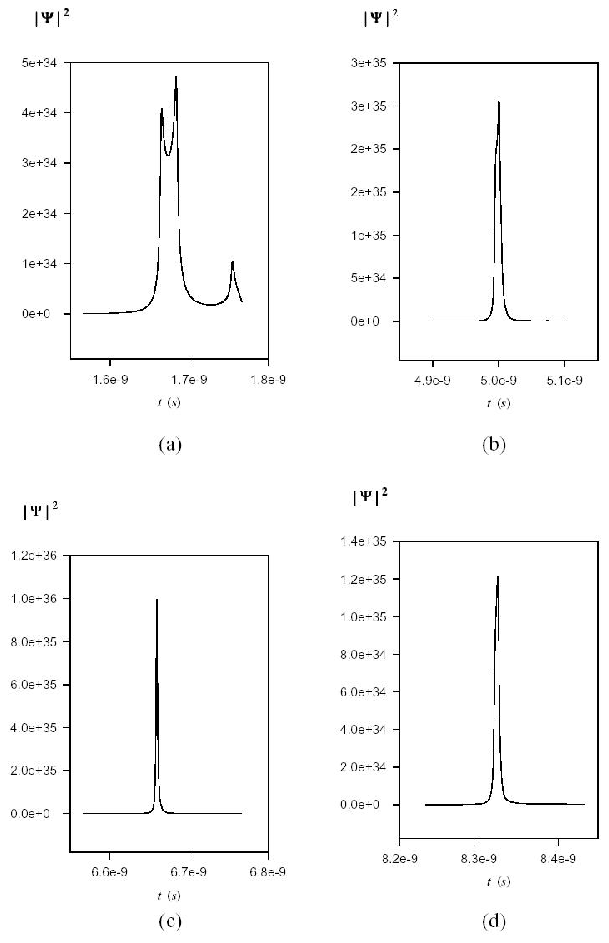}}
\end{center}
\caption{} \label{fig4}
\end{figure}

\newpage

\begin{figure}[!h]
\begin{center}
 \scalebox{1.4}{\includegraphics{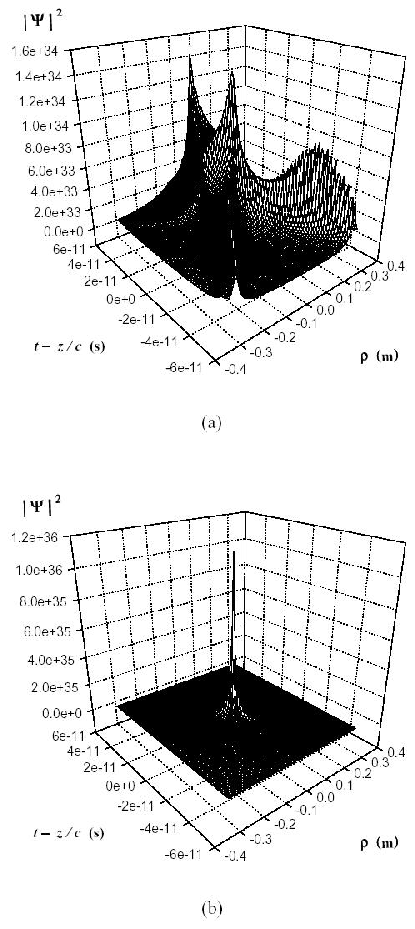}}
\end{center}
\caption{} \label{fig5}
\end{figure}

\newpage

\begin{figure}[!h]
\begin{center}
 \scalebox{1.4}{\includegraphics{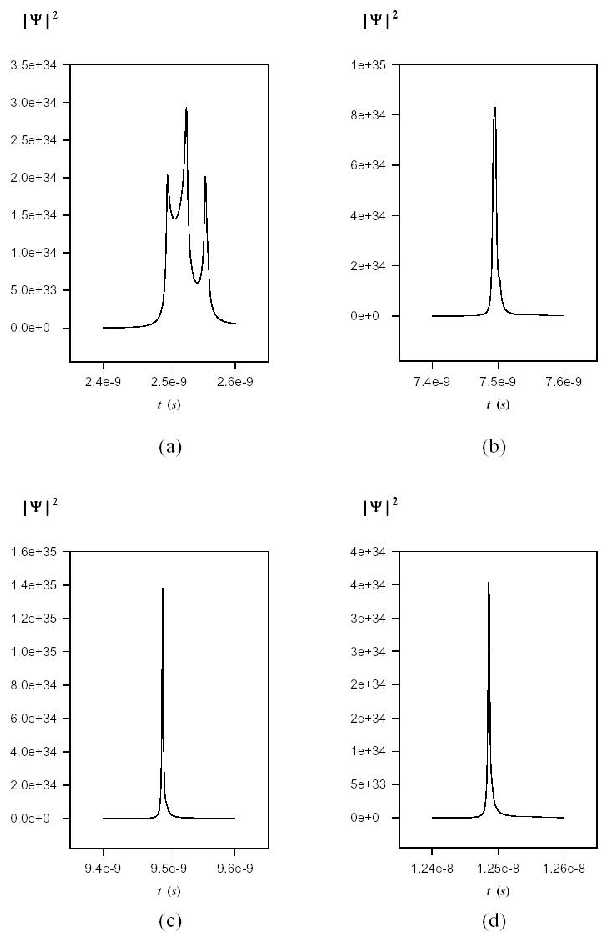}}
\end{center}
\caption{} \label{fig6}
\end{figure}

\

\h Although the influence of the size of the aperture has been
demonstrated only for the pulse given in Eq.(\ref{um}),
nevertheless one can easily extend the same analysis to
spatio-temporally focused pulses of other types [cf.
Eqs.(\ref{dois}),(\ref{tres}),(\ref{quatro})].

\

\

{\bf 7. -- Conclusions}\\

\h In conclusion, by a generalization of the discrete ``Temporal
Focusing" method[23], we have found new classes of Superluminal
waves. These new closed-form wave solutions show great potential
for space-time focusing. Indeed, we can tailor initially
spread-out pulses, so much so that they are strongly focused at a
point chosen a priori in space-time. In this work we have
demonstrated that these tailored pulses are easily adjustable by
varying the velocity spectra of their Superluminal wave
components.

\h We have also investigated the influence of having such pulses
launched from a finite-size aperture. It has been shown that,
because the individual Superluminal wave components, from which
the focusing pulse is synthesized, have aperture-dependent
diffraction-free lengths, the size of the aperture affects the
focusing position and magnitude. The described method is very
effective where all Superluminal wave components are propagating
within their diffraction-free range.

\

\

{\bf Acknowledgements}\\

\h The authors are very grateful to Hugo E.Hern\'andez-Figueroa
and K.Z.N\'obrega (FEEC, Unicamp), and to I.M.Besieris (Virginia
Polytechnic Institute) for continuous discussions and
collaboration. Useful discussions are moreover acknowledged with
T.F.Arecchi, A.M.Attiya and C.A.Dartora, as well as with
J.M.Madureira, S.Zamboni-Rached, V.Abate, F.Bassani, C.Becchi,
M.Brambilla, C.Cocca, R.Collina, R.Colombi, C.Conti, G.C.Costa,
G.Degli Antoni, L.C.Kretly, G.Kurizki, D.Mugnai, M.Pernici,
V.Petrillo, A.Ranfagni, A.Salanti, G.Salesi, J.W.Swart,
M.T.Vasconselos and
M.Villa.\\

\

\

{\bf 5. -- Figure Captions}\\

{\bf Fig.1 --} Space-time evolution of the Superluminal pulse
represented by Eq.(14); the parameter chosen values are
$a=10^{-12}\;$s; \ $V_\minrm = 1.001 \; c$; \ $V_\maxrm = 1.005 \;
c$ while the focusing point is at $z_\frm = 200\;$cm. \ One can
see that this solution is associated with a rather good
spatio-temporal focusing. \ The field amplitude at $z = z_\frm$
is 40.82 times larger than the initial one. \ The field amplitude
is normalized at the space-time point $\rho = 0, \ z = z_\frm, \
t = t_\frm$.

\

{\bf Fig.2 --} Space-time evolution of the Superluminal pulse
represented by Eq.(32). Now the parameters have the values $m =
1$, \ $a=10^{-12}\;$s (and therefore $\om_\crm = 1\;$GHz); \
$V_\minrm = 1.001 \; c$; \ $V_\maxrm = 1.005 \; c$); while the
focussing point is again at $z_\frm = 200\;$cm. \ Also this
solution is associated with a very good spatio-temporal focusing:
\ The field amplitude at $z = z_\frm$ is 1000 times higher than
the initial one. \ Once more, the field amplitude is normalized at
the space-time point $\rho = 0, \ z = z_\frm, \ t = t_\frm$.

\

{\bf Figs.3.} Surface plots of: \ (a) the initial excitation
$\big|\Psi\big|^2$ on the aperture plane $z'=0$, \ and \ (b) the
source-free pulse at the focusing point $z=z_\frm =200$ cm. The
spatio-temporally focused pulse corresponds to $a=10^{-12}$ s, \
$V_{min\,}=1.001\;c$ and $V_\maxrm =1.005\;c$. The radius of the
aperture is chosen equal to 20 cm.

\

{\bf Figs.4 --} The axial profiles of the field $\big|\Psi\big|^2$
radiated from a finite aperture at distances: \ (a) $z=50$; \ (b)
150; \ (c) 200 \ and \ (d) 250 cm. All other parameters are
chosen as in Figs.4.

\

{\bf Figs.5 --} Surface plots of: \ (a) the initial excitation
$\big|\Psi\big|^2$ on the aperture plane $z'=0$, \ and \ (b) the
source-free pulse at the \ focusing point $z=z_\frm =300$ cm. The
spatio-temporally focused pulse has $a=10^{-12}$ s, \
$V_{min\,}=1.001\;c$ and $V_\maxrm =1.005\;c$. The radius of the
aperture is chosen equal to 20 cm.

\

{\bf Figs.6 --} The axial profiles of the field $\big|\Psi\big|^2$
radiated from a finite aperture at distances: \ (a) $z=75$; \ (b)
225; \ (c) 285 \ and \ (d) 375 cm. All other parameters are
chosen as in Figs.6.

\newpage

\

{\bf Bibliography:}\hfill\break

[1] See, e.g., H.Bateman, {\em Electrical and Optical Wave
Motion} (Cambridge Univ.Press; Cambridge, 1915); \ J.A.Stratton,
{\em Electromagnetic Theory} (McGraw-Hill; New York, 1941),
p.356; \ R.Courant and D.Hilbert, {\em Methods of Mathematical
Physics} (J.Wiley; New York, 1966), vol.2, p.760.\hfill\break

[2] See, e.g., I.M.Besieris, A.M.Shaarawi and R.W.Ziolkowski, ``A
bi-directional traveling plane wave representation of exact
solutions of the scalar wave equation", {\em J. Math. Phys.},
vol.30, pp.1254-1269, June 1989; \ R.Donnelly and R.W.Ziolkowski,
``Designing localized waves", {\em Proc. Roy. Soc. London A},
vol.440, pp.541-565, March 1993.\hfill\break

[3] J.-y.Lu and J.F.Greenleaf, ``Nondiffracting X-waves: Exact
solutions to free-space scalar wave equation and their finite
aperture realizations", {\em IEEE Trans. Ultrason. Ferroelectr.
Freq. Control}, vol.39, pp.19-31, Jan.1992.\hfill\break

[4] E.Recami, ``On localized X-shaped Superluminal solutions to
Maxwell equations", {\em Physica A}, vol.252, pp.586-610,
Apr.1998; and refs. therein. \ For short review-articles, see, for
instance, E.Recami, ``Superluminal motions? A bird's-eye view of
the experimental situation", {\em Found. Phys.}, vol.31,
pp.1119-1135, July 2001; \ and Ref.[7].\hfill\break

[5] R.W.Ziolkowski, I.M.Besieris and A.M.Shaarawi, ``Aperture
realizations of exact solutions to homogeneous wave-equations",
{\em J. Opt. Soc. Am., A}, vol.10, pp.75-87, Jan.1993.\hfill\break

[6] M.Zamboni-Rached, E.Recami and H.E.Hern\'andez F., ``New
localized Superluminal solutions to the wave equations with
finite total energies and arbitrary frequencies", {\em Eur. Phys.
J., D}, vol.21, pp.217-228, Sept.2002.\hfill\break

[7] E.Recami, M.Z.Rached, K.Z.N\'obrega, C.A.Dartora \&
H.E.Hern\'andez F.: ``On the localized superluminal solutions to
the Maxwell equations", {\em IEEE Journal of Selected Topics in
Quantum Electronics}, vol.9, issue no.1, pp.59-73, Jan.-Feb.
2003.\hfill\break

[8] See, e.g., J.-y.Lu, H.-h.Zou and J.F.Greenleaf, ``Biomedical
ultrasound beam forming", {\em Ultrasound in Medicine and
Biology}, vol.20, pp.403-428, 1994.\hfill\break

[9] P.Saari and H.S\~{o}najalg, ``Pulsed Bessel beams", {\em Laser
Phys.}, vol.7, pp.32-39, Jan.1997.\hfill\break

[10] M.Zamboni-Rached, K.Z.N\'obrega, H.E.Hern\'{a}ndez-Figueroa
and E.Recami, ``Localized Superluminal solutions to the wave
equation in (vacuum or) dispersive media, for arbitrary
frequencies and with adjustable bandwidth" [e-print
physics/0209101], in press in {\em Opt. Commun.}\hfill\break

[11] C.Conti, S.Trillo, P.Di Trapani, G.Valiulis, A.Piskarskas,
O.Jedrkiewicz and J.Trull, ``Nonlinear electromagnetic X-waves",
{\em Phys. Rev. Lett.}, vol.90, paper no.170406, May 2003 [4
pages]; \ M.A.Porras, S.Trillo, C.Conti and P.Di Trapani,
``Paraxial envelope X-waves", {\em Opt. Lett.}, vol.28,
pp.1090-1092, July 2003. \
\hfill\break

[12] A. M. Attiya, {\em Transverse (TE) Electromagnetic X-waves:
Propagation, Scattering, Diffraction and Generation Problems},
Ph.D. Thesis, Cairo University, May 2001.\hfill\break

[13] See E.Recami, ``Classical tachyons and possible
applications", {\em Rivista N. Cim.}, vol.9(6), pp.1-178, 1986,
issue no.6, and refs. therein; \ A.O.Barut, G.D.Maccarrone and
E.Recami, ``On the shape of tachyons", {\em Nuovo Cimento A},
vol.71, pp.509-533, Oct.1982.\hfill\break

[14] J.Fagerholm, A.T.Friberg, J.Huttunen, D.P.Morgan and
M.M.Salomaa, ``Angular-spectrum representation of nondiffracting
X waves", {\em Phys. Rev., E}, vol.54, pp.4347-4352,
Oct.1996.\hfill\break

[15] J.-y.Lu and J.F.Greenleaf, ``Experimental verification of
nondiffracting X-waves", {\em IEEE Trans. Ultrason. Ferroelectr.
Freq. Control}, vol.39, pp.441-446, May 1992: In this case the
beam speed is larger than the {\em sound} (not of the light)
speed in the considered medium.\hfill\break

[16] P.Saari and K.Reivelt, ``Evidence of X-shaped
propagation-invariant localized light waves", {\em Phys. Rev.
Lett.}, vol.79, pp.4135-4138, Nov.1997.\hfill\break

[17] D.Mugnai, A.Ranfagni and R.Ruggeri, ``Observation of
superluminal behaviors in wave propagation", {\em Phys. Rev.
Lett.}, vol.84, pp.4830-4833, May 2000.\hfill\break

[18] P.Di Trapani, G.Valiulis, A.Piskarskas, O.Jedrkiewicz,
J.Trull, C.Conti and S.Trillo, ``Spontaneous formation of
nonspreading X-shaped wavepackets", e-print physics/0303083
(submitted for pub.).\hfill\break

[19] M.Z.Rached, E.Recami and F.Fontana, ``Superluminal localized
solutions to Maxwell equations propagating along a normal-sized
waveguide", {\em Phys. Rev., E}, vol.64, paper no.066603,
Dec.2001 [six pages]; \ M.Z.Rached, F.Fontana and E.Recami,
``Superluminal localized solutions to Maxwell equations
propagating along a waveguide: The finite-energy case", {\em
Phys. Rev., E}, vol.67, paper no.036620, March 2003 [seven
pages]; \ M.Z.Rached, K.Z.N\'obrega, E.Recami and H.E.Hernandez
F., ``Superluminal X-shaped beams propagating without distortion
along a coaxial guide", {\em Phys. Rev., E}, vol.66, 046617,
Oct.2002 [ten pages]; \ M.Zamboni-Rached and
H.E.Hern\'andez-Figueroa, ``A rigorous analysis of localized wave
propagation in optical fibers", {\em Opt. Commun.}, vol.191,
pp.49-54, May 2001.\hfill\break

[20] I.M.Besieris, M.Abdel-Rahman, A.Shaarawi and A.Chatzipetros,
``Two fundamental representations of localized pulse solutions of
the scalar wave equation", {\em Progress in Electromagnetic
Research (PIER)}, vol.19, pp.1-48, 1998.\hfill\break

[21]  S.He and J.y.Lu, ``Sidelobe reduction of
limited-diffraction beams with Chebyshev aperture apodization",
{\em J. Acoust. Soc. Am.}, vol.107, pp.3556-3559, June 2000; \
J.-y.Lu and S.He, ``High frame rate imaging with a small number
of array elements", {\em IEEE Trans. Ultrasound Ferroelec. Freq.
Control}, vol.46, pp.1416-1421, Nov.1999; \ J.-y.Lu,
``Experimental study of high frame rate imaging with
limited-diffraction beams", {\em IEEE Trans. Ultrasound
Ferroelec. Freq. Control}, vol.45, pp.84-97, Jan.1998; \ J.-y.Lu,
``Producing bowtie limited-diffraction beams with synthetic array
experiments", {\em IEEE Trans. Ultrasound Ferroelec. Freq.
Control}, vol.43, pp.893-900, Sep.1996; \ J.-y.Lu and
J.F.Greenleaf, ``Producing deep depth of field and
depth-independent resolution in NDE with limited-diffraction
beams", {\em Ultrasonic Imaging}, vol.15, pp.134-149,
1993.\hfill\break

[22]  A.A. Chatzipetros, A.M. Shaarawi, I.M. Besieris and M.
Abdel-Rahman, "Aperture synthesis of time-limited X-waves and
analysis of their propagation characteristics," {\em Journal of
the Acoustical Society of America}, vol. 103,  pp.2287-2295, May
1998; \ M. Abdel-Rahman, I.M. Besieris and A.M. Shaarawi, "A
comparative study on the reconstruction of localized pulses," in
{\em Proceedings of the IEEE Southeast Conference
(SOUTHEASTCON'97)}, pp.113-117 (Blacksburg, Virginia, April
1997).\hfill\break

[23] A.M.Shaarawi, I.M.Besieris and T.M.Said, ``Temporal focusing
by use of composite X-waves", {\em J. Opt. Soc. Am., A}, vol.20,
pp.1658-1665, Aug.2003.\hfill\break

[24] H.S\~{o}najalg, M.R\"{a}tsep and P.Saari, {\em Opt. Lett.},
vol.22, p.310, 1997.\hfill\break

[25] I.S.Gradshteyn and I.M.Ryzhik, {\em Integrals, Series and
Products}, 4th edition (Ac.Press; New York, 1965).\hfill\break

[26] A.T.Friberg, J.Fagerholm and M.M.Salomaa, ``Space-frequency
analysis of nondiffracting pulses", {\em Opt. Commun.}, vol.136,
pp.207-212, March 1997; \ J.Fagerholm, A.T.Friberg, J.Huttunen,
D.P.Morgan and M.M.Salomaa, ``Angular-spectrum representation of
nondiffracting X waves", {\em Phys. Rev., E}, vol.54,
pp.4347-4352, Oct.1996; \ P.Saari, in {\it Time's Arrows, Quantum
Measurements and Superluminal Behavior}, D.Mugnai et al. editors
(C.N.R.; Rome, 2001), pp.37-48.\hfill\break

[27] D.Mugnai, A.Ranfagni and R.Ruggeri, ``Pupils with
super-resolution", {\em Phys. Lett., A}, vol.31, pp.77-81, March
2003; \ G.Toraldo di Francia, {\em Supplem. Nuovo Cimento},
vol.9, pag.426, 1952.

\end{document}